\documentclass[conference]{IEEEtran}
\IEEEoverridecommandlockouts
\usepackage{cite}
\usepackage{amsmath,amssymb,amsfonts}
\usepackage{algorithmic}
\usepackage{hyperref}
\usepackage{graphicx}
\usepackage{textcomp}
\usepackage{xcolor}
\def\BibTeX{{\rm B\kern-.05em{\sc i\kern-.025em b}\kern-.08em
    T\kern-.1667em\lower.7ex\hbox{E}\kern-.125emX}}
\begin{document}

\title{\huge Localization using Mobile Wireless Sensor Networks}

\author{\IEEEauthorblockN{Aaron John-Sabu}
\IEEEauthorblockA{\textit{Dept. of Electrical Engineering} \\
\textit{Indian Institute of Technology Bombay}\\
Mumbai, India\\
aaronjs@iitb.ac.in}
}

\maketitle

\begin{abstract}
Wireless Sensor Networks (WSNs) are groups of spatially distributed and dedicated autonomous sensors for monitoring (and recording) the physical conditions of the environment (and organizing the collected data at a central location). They have been a topic of interest due to their versatility and diverse capabilities despite having simple sensors measuring local quantities such as temperature, pH, or pressure. We delve into understanding how such networks can be utilized for localization, and propose a technique for improving conditions of living for animals and humans on the IIT Bombay campus.
\end{abstract}


\section{INTRODUCTION}
In a society that is booming with collaborative and wide-spread activities, both human-based and autonomous, it is a vital requirement to incorporate similar networks that can monitor and support human needs such as pollution control, navigation, industrial monitoring etc. Utilizing wireless sensors networks for such functionalities further helps in the reduction of power consumption, work time, and difficulties in implementation. We study the work performed on the application of WSNs in localization. The learnings are conglomerated to develop a mechanism for the maintenance of wildlife in the IIT Bombay campus, particularly dogs.

\section{PREVIOUS RESEARCH}
In recent times, the incorporation of WSNs into populated societies have improved conditions of life and work, from increasing crop yield and optimizing utility distribution systems (electricity, water, etc.) to reducing wasted power and avoiding traffic jams. \cite{a1} elucidates how WSNs have developed throughout the years, from more expensive and fewer sensors to cheaper but larger quantities of sensors, from discrete circuits and multi-chip solutions to system-on-chip (SoC) devices, and from one-way communication links to bidirectional links and mesh and star designs. These optimizations have led to widespread applications of WSNs one of them being navigational localization of objects and bodies using methods such as trilateration.
\subsection{Localization using Wireless Sensor Networks}
Localization can be performed using multiple methods, a very common one being GPS. However, the accuracy of GPS data is relatively low ($>10m$). \cite{b1} proposes a more accurate localization method wherein a wireless sensor network of ZigBees is developed and their relative signal strengths are used for trilateration-based localization. The Mobile Target Tracking (MTT) problem intends to find the moving path of a target in a field based on target locations that are sampled at random intervals. \cite{b2} discusses several algorithms for solving this problem using two aspects: determining the current location of the target (localization, path tracing), and processing information collaboratively among multiple sensor nodes. Traditional methods involving the informed selection of sensors, binary sensor-based methods with centralized and distributed architectures, and other methods based on triangulation are suggested for tracking, while information can be processed using leader-based algorithms or distributed algorithms.\\
Unlike open environments, locations with several obstructions or jamming hinder the proper function of the Global Position System (GPS). In such scenarios, it is necessary to develop a positioning system that can complement GPS. \cite{b3} proposes a pedestrian navigation system (PNS) wherein heterogeneous agents are used for sensing, communication and relaying information. Here, sensors such as accelerometers and magnetometers are used as part of a dead reckoning (odometry) mechanism to localize the user. The NavMote (sensor on the user) exchange information with NetMotes (predetermined sensors) when they are in range; otherwise it works on its own. This system has been tested in both indoor and outdoor environments, giving satisfactory performance at a distance accuracy of $\pm1\%$ and a heading accuracy of $1^\circ$.\\
Localization for mobile targets can be performed using beacon-based methods, wherein some beacons aware of their positions provide geographic information to ordinary sensor nodes to localize, whose precision increases with the number of beacons. \cite{b4} proposes an algorithm utilizing mobile beacons that traverse the network deployment area and broadcast beacon packets to generate a number of virtual beacons. The distance between the sensor node and the beacon can be calculated using RSSI (Received Signal Strength Indicator). Since a single-mobile-beacon system can have issues such as co-linearity due to the straight line moving trajectory of the mobile beacon, a three-mobile-beacon-assisted mechanism is suggested wherein the object (sensor node, $S_i$) is localized as the weighted centroid of the three beacon positions: $$ S_i(\hat{x}_{si},\hat{y}_{si})=\frac{\sum_{j=1}^mw_{ij}V_j(x_{vj},y_{vj})}{\sum_{j=1}^mw_{ij}}, w_{ij}=\frac{1}{(d_{ij})^g},\label{eq} $$
where $V_j$ is the $j$th virtual beacon, $d_{ij}$ is the distance between $S_i$ and $V_j$, and $g$ is an adjustable degree.\\
While anchors in anchor node-based schemes acquire their positions in advance using GPS systems or artificial arrangement to locate unknown nodes, unknown nodes in anchor node-free schemes are located using the connectivity information between unknown nodes and anchor nodes. The former achieves better localization accuracy but require a large number of anchor nodes which increases the energy consumption and hardware cost of WSNs. \cite{b5} proposes using mobile anchor nodes to maximize the localization accuracy while decreasing the energy consumption of WSNs. An anchor node moves based on an equilateral triangle trajectory in a WSN area and broadcasts position and time messages periodically which, on reception, are used via RSSI-based trilateration to determine the position of unknown nodes. Though sensitive to the standard deviation of noise, the algorithm reduces the number of beacon positions, trajectory lengths and node density. Moreover, it remains robust at high traveling speeds of the anchor node.

\subsection{Optimization of WSNs and Other Applications}
The distributed detection process involves sensor nodes that are deployed randomly in a field to collect data from the surrounding environment. \cite{c1} describes and compares three detection schemes: centralized, distributed, and quantized, based on parameters such as detection probability and overall probability of error, measured against varying energy consumption, time periods and distances from the sensors to the command center. The robustness of each scheme is also tested using node destruction and partial observation deletion. The distributed scheme is observed to be superior in energy consumption (especially over large distances) and robustness. Although the centralized scheme uses fewer nodes, the distributed option needs fewer than twice that number to achieve the same detection performance.\\
The neighbor discovery problem deals with situations where sensor nodes find their neighbors constantly in mobile sensor networks by communicating with each other while in motion and forward the collected information to a central command center. Here, the active and dormant status of sensors can be controlled, hence reducing the energy consumption significantly. However, this may cause additional discovery latency since discovery is possible only when neighboring nodes have overlapping active slots. Previous research had introduced the group-based method where a third state for waking up actively is used to communicate the schedule and verify the neighborhood of nodes. \cite{c2} proposes an algorithm that considers the embedded spatial properties and actively modifies the active time of nodes depending on the number of undiscovered neighbors. This has been tested using simulations and the discovery time has been found to be minimal when compared to algorithms presented in existing literature.\\
\cite{c3} proposes a wireless sensor network distributed in the physical environment for sensing, actuation and communication such that indoor lighting is automated, while also giving override capabilities to users based on individual preferences. The system is decomposed into smaller pseudo-static subnets and issues such as nodes belonging to multiple subnets are solved using a multi-agent system approach wherein agent actions are averaged or voting can be used to determine the action. Furthermore, a supervised learning model can be used to predict the action of the user and act accordingly while remaining bounded to physical constraints. This can be replaced by model-free reinforcement learning algorithms so that the agents are trained when a model is unavailable or too difficult to compute in a complicated real-world environment.

\section{The Effect of TAG LINKAGES for MOBILE WSN-based LOCALIZATION}
\subsection{Introduction}
The Indian Institute of Technology Bombay campus, spread across over 500 acres, is home not only to humans but also a wide array of plants and animals from leopards and crocodiles to cows, dogs and cats. Unfortunately, in recent times, human activities in the institute have been disrupted by stray dogs. Also, survival for these dogs has also become difficult due to the movement of heavy-duty construction vehicles along the same routes posing a threat to their lives. This report proposes a technique to track and guide dogs while avoiding harm to the human population as well as the dogs in the institute.
\subsection{Procedure and Solution}
Based on the concept of mobile target tracking as discussed in \cite{b2}, a network of distributed sensors may be placed at suitable locations in the institute. However, owing to the large size of the campus, this is not a scalable solution since the number of sensors and the corresponding energy consumption will be huge. Moreover, the distribution of dogs across the institute need not be uniform, due to which placing sensors at certain locations will not be efficient although there is a possibility for a small number of dogs to visit these areas.\\
Incorporating ideas from \cite{b5}, the movable sensors on the dogs can behave as anchors using which other movable sensors can localize themselves via trilateration. The number of fixed sensors can hence be reduced if the number of dogs is large. This is demonstrated using a simulation wherein the number of fixed sensors has been reduced to $6$ in a $10m\times10m$ area where each sensor can sense sensors within a radius of $7m$. Three movable sensors are placed such that the size of their point represents the uncertainty of finding their location.\\
We define the neighbors of a sensor at a particular moment to be the sensors capable of connecting to and providing information about position and time to the sensor for trilateration at that instant. A simple geometric uncertainty principle is used wherein the uncertainty of the location of the sensor increases given the number of neighbors is less than $4$ (with a lower rate for $3$ neighbors) and decreases otherwise at an exponential rate proportional to the number of neighbors.\\
Free-ranging dogs generally exhibit territoriality as studied in \cite{d1} and those in the institute have been observed to do so in considerably small pieces of land such as individual hostels and small strips of roads. Similarly, the three sensors in the simulation are constrained within particular regions.
\subsection{Results}
\begin{figure}[thpb]
    \centering
    \framebox{
      \parbox{2.5in}{\includegraphics[width=0.95\linewidth]{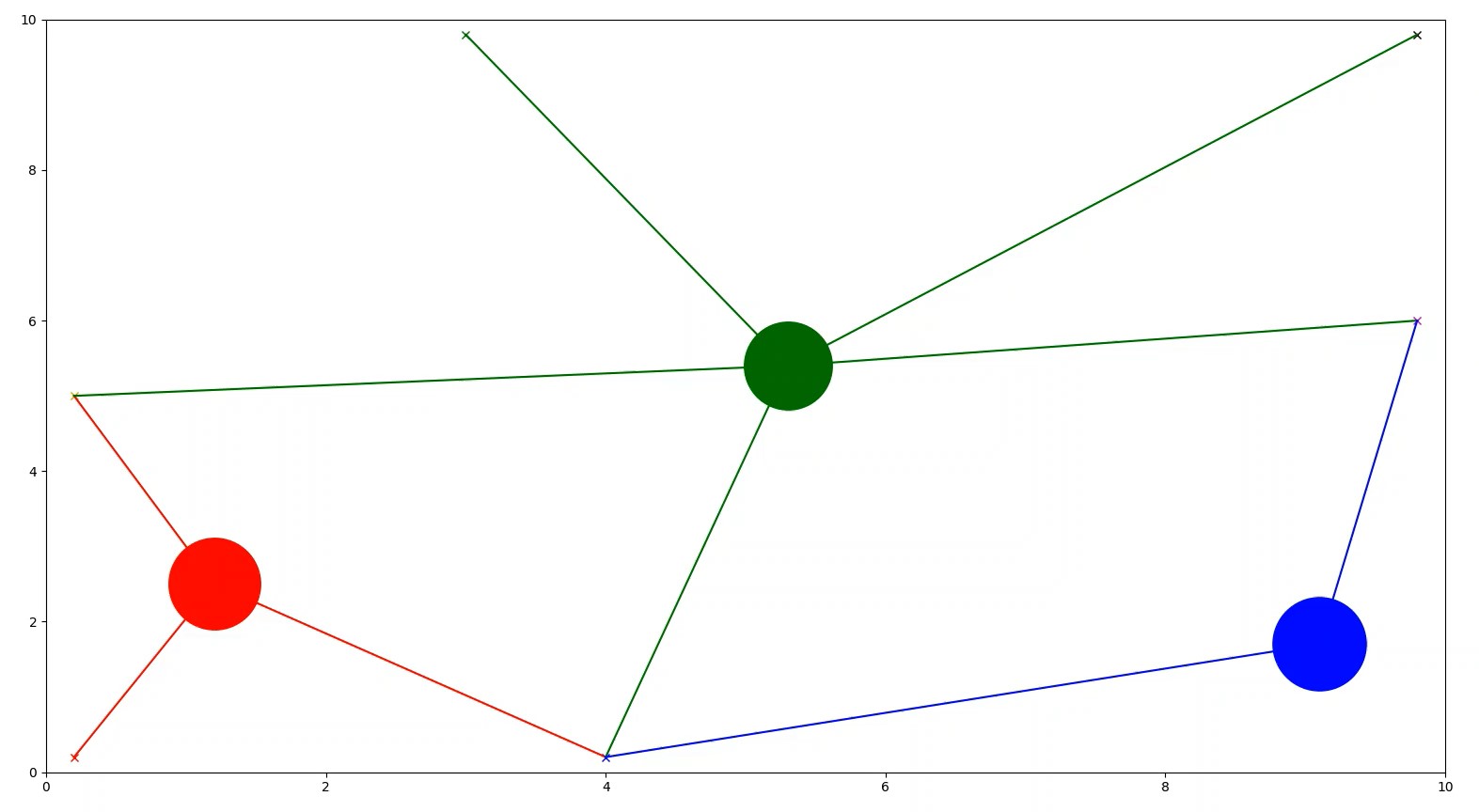}}
    }
    \caption{Initial Configuration - Without Tag Links}
    \label{NoStart}
\end{figure}

\begin{figure}[thpb]
    \centering
    \framebox{
      \parbox{2.5in}{\includegraphics[width=0.95\linewidth]{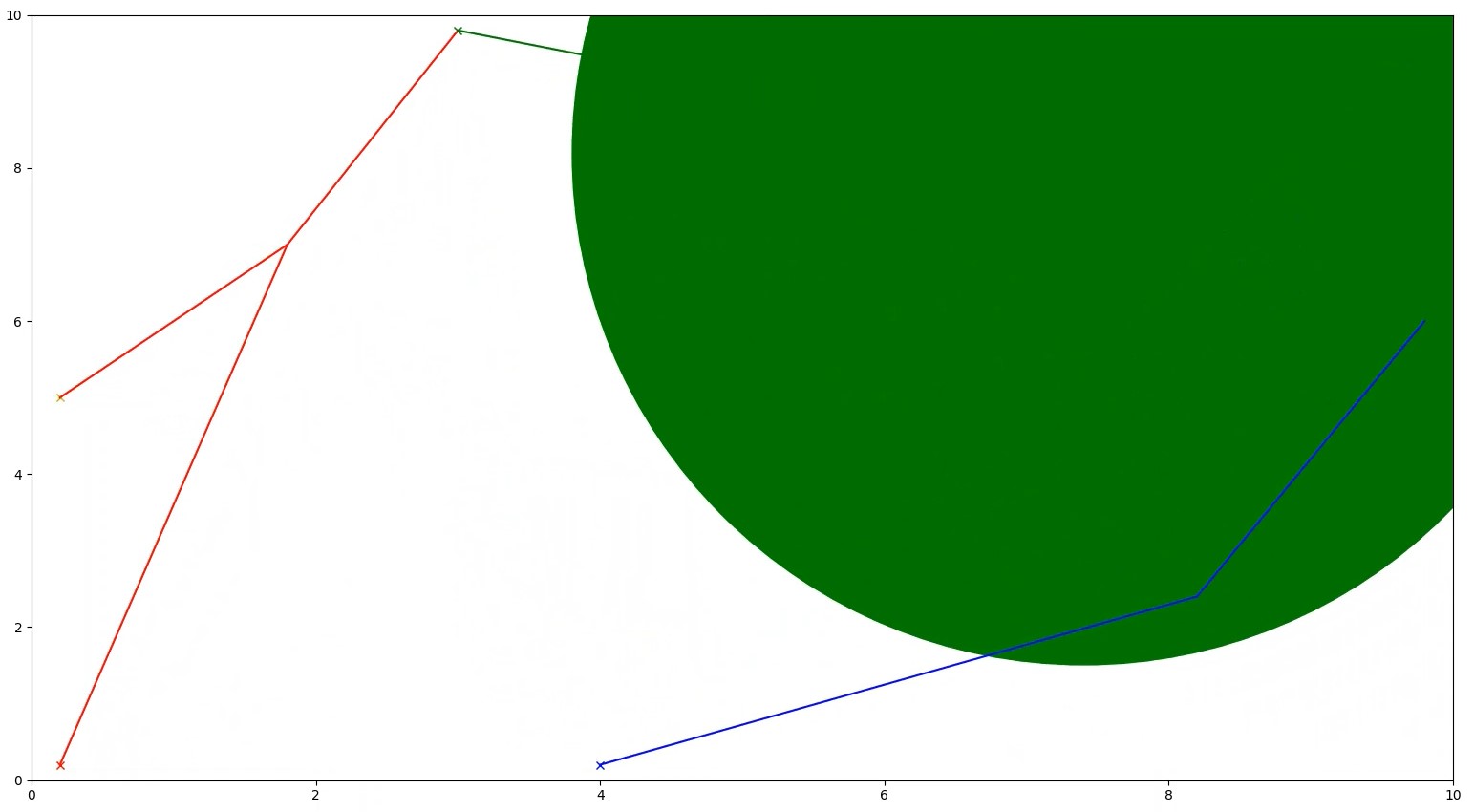}}
    }
    \caption{Final Configuration - Without Tag Links}
    \label{NoTagLinks}
\end{figure}

\begin{figure}[thpb]
    \centering
    \framebox{
      \parbox{2.5in}{\includegraphics[width=0.95\linewidth]{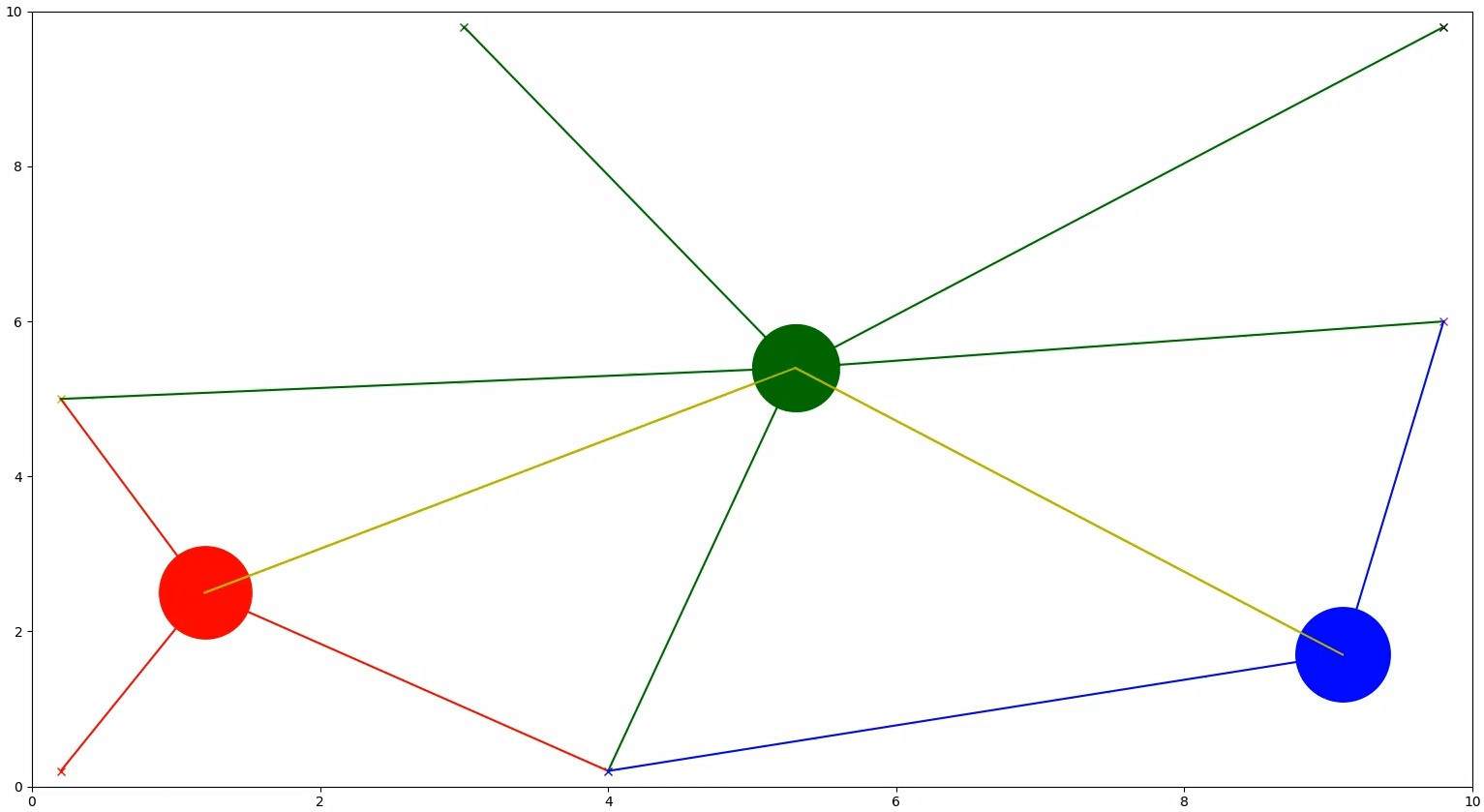}}
    }
    \caption{Initial Configuration - With Tag Links}
    \label{Start}
\end{figure}

\begin{figure}[thpb]
    \centering
    \framebox{
      \parbox{2.5in}{\includegraphics[width=0.95\linewidth]{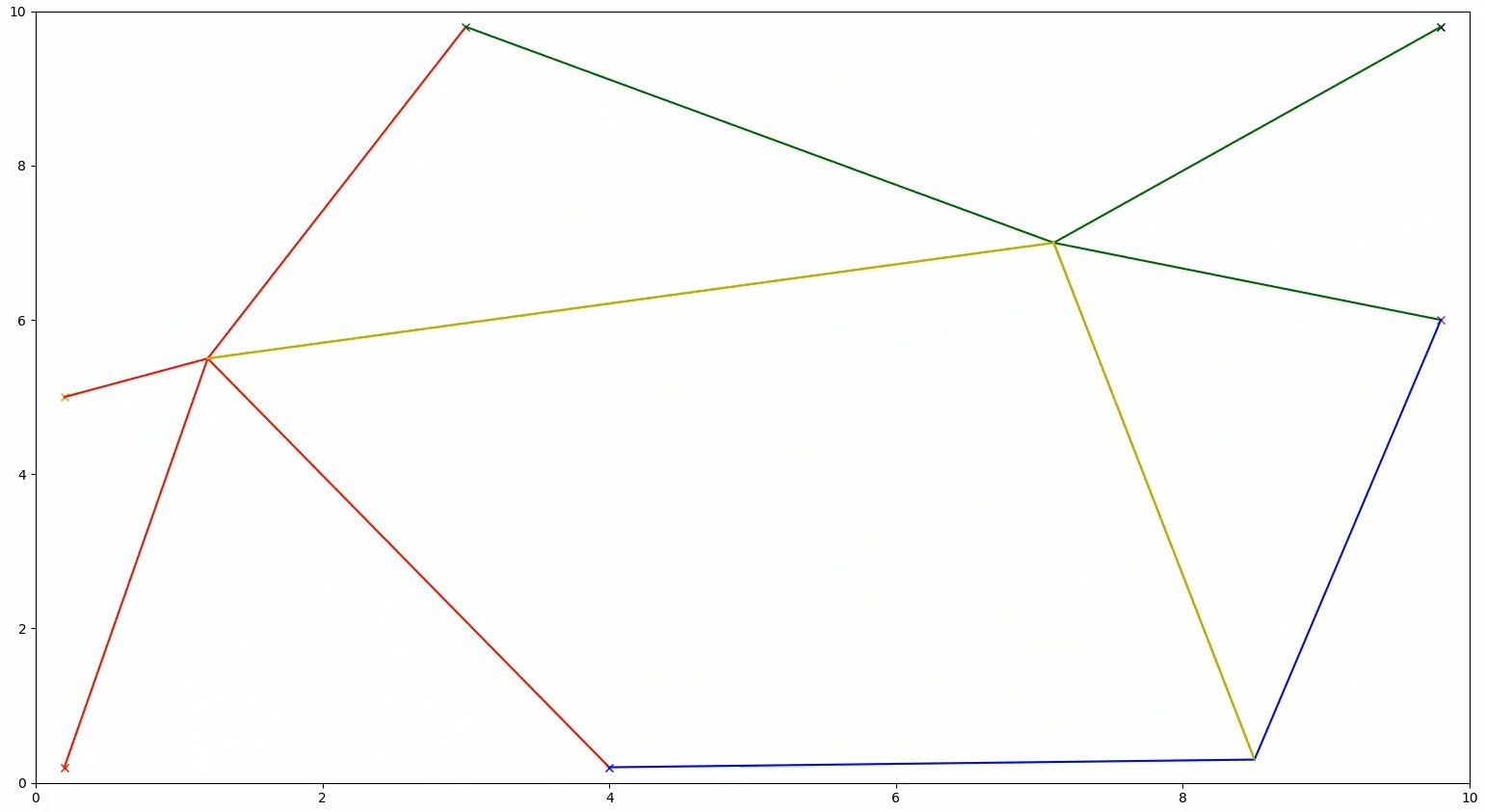}}
    }
    \caption{Final Configuration - With Tag Links}
    \label{TagLinks}
\end{figure}

The mechanism is simulated in two-dimensional space within the given dimensions. We initialize the sensors with some positional uncertainty as shown in Fig. \ref{NoStart}.
On deactivating links between tags (i.e., they do not behave as mutual anchors), there are larger periods when the number of links is $2$ due to which the uncertainty of some sensors is observed to diverge with time as depicted in Fig. \ref{NoTagLinks}.\\
The second simulation involves links between neighboring tags. This starts from the beginning of the simulation as shown in Fig. \ref{Start}.
For simplicity, we assume that the range of the tag sensors is equal to that of the anchors although, in a practical scenario, this might not be the case since the former will consist of simpler circuits and transceivers, hence resulting in smaller ranges. It is observed from Fig. \ref{TagLinks} that the uncertainty of the previously mentioned sensors does not diverge since the sensors are in contact with each other for several periods which previously had led to the divergence of uncertainty.
\section{FUTURE SCOPE}
Wearing traditional sensors for a long duration may be detrimental to the health of the dog, and the sensing quality may degrade from the environmental and hygienic conditions. Biosensors and flexible electronics may be more appropriate long-term options. The long-term effects of biologic cybernetics and electronic stimulation on dogs is open to research although there has been progress in short-duration studies on several animal species as shown in \cite{d2}, \cite{d3} and \cite{d4}.\\~\\
The simulation code related to this project is available at  \href{https://github.com/aaronjohnsabu1999/Localization-Using-WSNs}{github.com:aaronjohnsabu1999/Localization-Using-WSNs}
\section{ACKNOWLEDGMENTS}
I am grateful to Prof. Siddharth Tallur for the guidance and support throughout this project as the course instructor for EE 617 (Sensors in Instrumentation) at IIT Bombay.



\begin{thebibliography}{00}
\bibitem{a1} A. Benefit, ``The evolution of wireless sensor networks'', [online] Available: https://www.silabs.com/documents/public/white-papers/evolution-of-wireless-sensor-networks.pdf
\bibitem{b1} Lee, J-H, Kim, K, Lee, S-C, and Shin, B-S, ``An Efficient Localization Method Based on Adaptive Optimal Sensor Placement,'' Int. Journal of Distributed Sensor Networks, 2014
\bibitem{b2} Gupta, A, Gui, C, and Mohapatra, P, ``Mobile Target Tracking using Sensor Networks,'' Mobile, Wireless, and Sensor Networks: Technology, Applications, and Future Directions, 2006
\bibitem{b3} Fang, L, Antsaklis, PJ, et al., ``Design of a Wireless Assisted Pedestrian Dead Reckoning System—The NavMote Experience,'' IEEE Transactions on Instrumentation and Measurement, 2016
\bibitem{b4} Cui, H-q, Wang, Y-l, et al, ``Three-Mobile-Beacon Assisted Weighted Centroid Localization Method in Wireless Sensor Networks,'' Int. Conf. on Software Engineering and Service Science, 2011
\bibitem{b5} Jiang, J, Han, G, Xu, et al., ``LMAT: Localization with a Mobile Anchor Node Based on Trilateration in Wireless Sensor Networks,'' IEEE Global Telecommunications Conf., 2011
\bibitem{c1} Yu, L and Ephremides, A, ``Detection, Energy, and Robustness in Wireless Sensor Networks,'' Mobile, Wireless, and Sensor Networks: Technology, Applications, and Future Directions, 2006
\bibitem{c2} Niu, Q, Bao, W, and Xia, S, ``An Improved Group-Based Neighbor Discovery Algorithm for Mobile Sensor Networks,'' Int. Journal of Distributed Sensor Networks, 2014
\bibitem{c3} Sandhu, JS, Agogino, AM, and Agogino AK, ``Wireless sensor networks for commercial lighting control: Decision making with multi-agent systems,'' AAAI Workshop on Sensor Networks, 2004
\bibitem{d1} Majumder, SS, et al., ``A dog’s day with humans – time activity budget of free-ranging dogs in India,'' Current Sciences 10(6), 2014
\bibitem{d2} Johnson, LA, and Fuglevand, AJ, ``Mimicking muscle activity with electrical stimulation,'' Journal of Neural Engineering vol. 8(1), 2011
\bibitem{d3} Rezaee, Z, and Kobravi, HR, ``Human Gait Control Using Functional Electrical Stimulation Based on Controlling the Shank Dynamics,'' Basic and Clinical Neuroscience vol. 11(1), 2020
\bibitem{d4} Cao, F, Zhang, C, Vo Doan, TT, et al, ``A Biological Micro Actuator: Graded and Closed-loop Control of Insect Leg Motion by Electrical Stimulation of Muscles,'' PLOS ONE 9(8), 2014
\end{thebibliography}
\end{document}